# Inelastic electron tunneling spectroscopy in molecular junctions: Peaks and dips

Michael Galperin and Mark A. Ratner

Department of Chemistry, Northwestern University, Evanston IL, 60208

and

Abraham Nitzan

School of Chemistry, The Sackler Faculty of Science, Tel Aviv University, Tel Aviv, 69978, Israel

## Abstract

We study inelastic electron tunneling through a molecular junction using the non-equilibrium Green function (NEGF) formalism. The effect of the mutual influence between the phonon and the electron subsystems on the electron tunneling process is considered within a general self-consistent scheme. Results of this calculation are compared to those obtained from the simpler Born approximation and the simplest perturbation theory approaches, and some shortcomings of the latter are pointed out. The self-consistent calculation allows also for evaluating other related quantities such as the power loss during electron conduction. Regarding the inelastic spectrum, two types of inelastic contributions are discussed. Features associated with real and virtual energy transfer to phonons are usually observed in the second derivative of the current $I$ with respect to the voltage $\Phi$ when plotted against $\Phi$. Signatures of resonant tunneling driven by an intermediate molecular ion appear as peaks in the first derivative $dI/d\Phi$ and may show phonon sidebands. The dependence of the observed vibrationally induced lineshapes on the junction characteristics, and the linewidth associated with these features are also discussed.

# 1. Introduction

Experiments on conduction in molecular wire junctions are becoming more common, as the community seeks to understand the principles that govern how electrical charge can be carried by individual molecules.[1] Early experiments focused on the absolute conduction and on trends such as dependence on wire length, molecular structure and temperature.[2] An intriguing issue is the role played by nuclear motions in the conduction process. This issue is of interest on several accounts. First, it underlines the interplay between coherent transport by carrier tunneling and/or band motion, polaronic conduction and incoherent, thermally assisted hopping transport.[3] Indeed, the importance of the full hopping regime, in which charges are definitely localized on the molecular bridge, has been demonstrated both in the Coulomb blockade limit[4] and in a polaron-type localization situation.[5] Secondly, it is directly relevant to the issue of junction heating.[6,7] Also, vibronic interactions accompanying electron transport may lead to specific nuclear motions such as rotations,[8,9] lateral hopping of molecules on the surface,[10] atomic rearrangements[11] and chemical reactions.[12] Finally, nuclear motions can directly manifest themselves as inelastic signals in the current-voltage spectra. Inelastic electron tunneling spectroscopy (IETS) has been an important tool for identifying molecular species in tunnel junctions for a long time.[13] With the development and advances in scanning tunneling microscopy (STM) and spectroscopy (STS) it has proven invaluable as a tool for identifying and characterizing molecular species within the conduction region.[9,14-19] Indeed, this is the only direct way to ascertain that a molecular species indeed participates in the conduction process, and at the same time to provide important spectroscopic and structural data on the conducting molecule, in particular information on the strength of the vibronic coupling itself.

The interpretation of electronic transport in molecular junctions has so far been done largely in the context of an elastic scattering picture ultimately derived from the Landauer approach.[20] Inelastic conductance through models of molecular wires was considered theoretically by several workers. Ness and Fisher[21] and Todorov and coworkers[22] have considered the inelastic propagation of an electron through the junction as a multichannel scattering problem, while Segal, Nitzan, Ratner and coworkers[6,23] as well as May, Petrov and coworkers[24] have used formalisms based on generalized quantum master equations. The influence of the contact population (Pauli principle) on the inelastic process is disregarded in these works as is the influence of the electronic subsystem on the phonon dynamics.

A systematic framework describing transport phenomena in interacting particle systems is based on the nonequilibrium Green's function (NEGF) formulation.[25] For the particular case of IETS, while simple perturbative treatments using Herzberg-Teller like analysis of the molecular Green's function or the electron propagator[26] may be useful for rough estimates using realistic molecular models, such a heuristic approach is not fully consistent with the non-equilibrium conditions under which such measurements are done as well as with the boundary restrictions imposed by the Pauli principle. Also, by using the lowest order in the electron-phonon interaction it misses important interference effects and cannot deal with the mutual effects of electron correlation and vibronic coupling. The NEGF formalism can readily handle such problems,[27,28] though its complexity may limit its usefulness to relatively simple molecular models. Several recent discussions of inelastic transport in microscopic junctions are particularly relevant to the present discussion[29-32] (see also the early treatment by Caroli et al[27]).

Such an approach was recently taken by Ueba and coworkers[30-32] who have applied the NEGF formalism to the resonant level model of phonon assisted tunneling where a single bridge level represents a junction connecting two free electron reservoirs while being also coupled to a single harmonic mode. The free particle Hamiltonian is

$$\hat{H}_0 = E_1 \hat{c}_1^\dagger \hat{c}_1 + \sum_{k \in L,R} \varepsilon_k \hat{d}_k^\dagger \hat{d}_k + \Omega_0 \hat{a}^\dagger \hat{a} \quad (1)$$

where $\hat{c}_1^\dagger$ and $\hat{c}_1$ are creation and annihilation operators for electrons on the bridging level of energy $E_1$, $\{k\} = \{l\}, \{r\}$ are sets of electronic states representing the left (L) and the right (R) electrodes with the corresponding creation and annihilation operators $\hat{d}_k^\dagger$ and $\hat{d}_k$ and $\hat{a}^\dagger$ and $\hat{a}$ are creation and annihilation operator for the phonon mode of frequency $\Omega_0$. The interactions are given by

$$\hat{H}_1 = \sum_{k \in L,R} \left( V_{k1} \hat{d}_k^\dagger \hat{c}_1 + h.c. \right) + M \left( \hat{a}^\dagger + \hat{a} \right) \hat{c}_1^\dagger \hat{c}_1 \quad (2)$$

Within this mode Ueba et al have reproduced and improved results obtained earlier by Persson and Baratoff[33-35] where inelastic tunneling spectra were analyzed in the leading order $M^2$ of the electron phonon interaction. Persson and Baratoff have observed (following Davis[36]) that in this order there is an important correction to the elastic component of the tunneling current at the onset ($|e\Phi| = \hbar\Omega_0$ where $\Phi$ is the bias potential) of the inelastic channel. This contribution to the tunneling flux stems from what may be seen as interference between purely elastic current amplitude that does not involve electron-phonon interaction and the elastic amplitude associated with two electron-phonon interaction events involving virtual phonon emission and absorption. Similarly, Lorente and M.Persson[29] have recently generalized the Tersoff-Hamann approach to the tunneling in STM junctions, using many-body density functional theory in conjunction with the NEGF formulation of Caroli et al[27] to obtain the change in the local density of substrate electronic states caused by a weak electron-phonon coupling. This approach again yields the inelastic contribution to the tunneling flux in the lowest order in the electron-phonon coupling, as well as the elastic correction of the same order. This formalism was later applied to formulate symmetry propensity rules for vibrationally inelastic tunneling.[15]

At the threshold $|e\Phi| = \hbar\Omega_0$ of the inelastic tunneling channel both the elastic and the inelastic fluxes change, with the latter obviously increasing from its zero value below threshold. In contrast, as first noted by Person and Baratoff,[35] depending on the energetic parameters of the system, the correction to the elastic current may be negative. Furthermore, this negative change in the elastic tunneling component may outweigh the positive contribution of the inelastic current, leading to a negative peak in the second derivative of the current/voltage relationship. Such negative features have indeed been observed in single-molecule vibrational spectroscopy of methyl isocyanide adsorbed on aluminia-supported rhodium particles[37] and of oxygen molecules chemisorbed on Ag(100).[16] It should be noted that not only the sign but also the shape of these peaks depend on the energetic parameters of the system,[38] and recent results by Reed and coworkers[18] that show relatively strong derivative-like features in the low temperature IETS spectrum of C8 alkane thiols may be another manifestation of the same effect.

The Person-Baratoff analysis[33-35] as well as its reassessment by Ueba and coworkers[30-32] correspond to the limit where electron transmission across the junction is a low probability event that does not disturb the electron distribution in the leads. In the opposite limit where the leads-bridge coupling is strong so that the transmission probability is nearly 1 (and the single channel conduction given by $\sim e^2/\pi\hbar$) we may encounter the situation where in the negatively biased lead backscattered electrons of energies in the conduction window between the left ($E_{FL}$) and right ($E_{FR}$) Fermi energies are locally depleted near the junction. In this case the onset of inelastic scattering at $\hbar\omega_0 = |e\Phi|$ can give rise to increased reflection (which would otherwise be impossible by the Fermi exclusion) and, consequently, a negative step in the conduction. This is presumably the dominant mechanism for observed negative peaks in $d^2I/d\Phi^2$ in point contact spectroscopy characterized by large transmission probabilities.[19,39]

The inelastic tunneling features discussed so far are usually observed as peaks (or dips) in the second derivative of the current voltage relationship at the threshold where the electronic energy associated with the bias voltage $e\Phi$ just matches the oscillator energy $\hbar\Omega_0$. Another manifestation of electron phonon interaction in inelastic tunneling can be observed as phonon sidebands of resonance tunneling features.[17] Figure 1 illustrates the energetic scenarios for these different processes at *T*=0. The shaded areas on the right and left denote the continuous manifolds of states of the two leads where the lines separating the occupied and unoccupied states are the corresponding Fermi energies. In this zero order picture a wavefunction of the overall system is a product of lead wavefunctions and a molecular state. One of the two manifolds shown for the right lead represents the ground vibrational state of the molecule. The other (diagonally shaded) corresponds to the molecule in the first excited vibrational state. We assume for simplicity that a potential bias $\Phi$ amounts to raising the energy of the left lead states without affecting the other states of the model and that conduction takes place via the LUMO (lowest unoccupied molecular orbital) state of the bridge molecule, i.e. resonance tunneling occurs via negative ion states. As the bias is increased the first type of vibrational feature is obtained when $e\Phi = \hbar\Omega_0$ and is seen as a peak (dip) in the $d^2I/d\Phi^2$ vs. $\Phi$ spectrum. The second type appears when $\Phi$ is large enough so that $\mu_L$ just exceeds $E_1$: resonance tunneling features appear as peaks in the conductance ($dI/d\Phi$ vs. $\Phi$) spectrum. At low temperature the first such peak corresponds to the ground state of the molecular negative ion state, however if the width associated with the coupling of this state to the metal is small enough additional peaks are expected when $\mu_L$ traverses $E_1 + n\hbar\Omega_0$, $n = 1, 2,...$ (peaks associated with negatives values of *n* are also possible at higher temperatures). Similar features can be seen in the conduction dependence on a gate potential which changes the position of $E_1$ relative to the Fermi energies. The process is analogous to resonance Raman scattering[40] and both the proximity to resonance and the fact that the transient electronic state is charged imply strong vibronic coupling and consequently long progression of vibrational satellites, as indeed seen in the experimental results of Refs. [17]. We are not familiar with a theoretical treatment of these observations, however the same resonant tunneling model described above, when properly generalized to allow for strong electron-phonon coupling should provide a suitable framework.

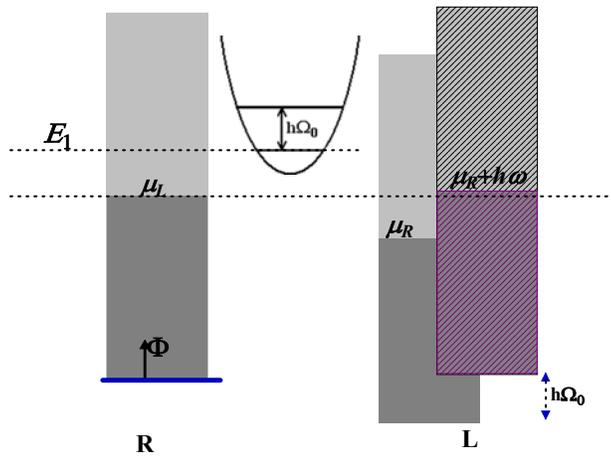

Fig. 1. A schematic view of the level structure for inelastic electron tunneling. The shaded areas on the right and left denote the continuous manifolds of states of the two leads where the lines separating the occupied and unoccupied states are the corresponding Fermi energies. The parabola represents the nuclear potential surface of the bridge. For the right lead two manifolds are shown: one where the corresponding molecular state is the ground vibrational state of the molecule, and the other (diagonally shaded) where the molecule is in the first excited vibrational state. The horizontal dotted lines at heights $\mu_L$ and $E_1$ are added to guide the eye.

In this paper we generalize the NEGF approach to inelastic tunneling spectroscopy in several ways. First, many electronic states and phonon modes are taken into account, allowing the usual treatments of a real molecular bridge in any convenient representation. Secondly, the vibrational modes are taken as two subsets. The primary subset are those modes that directly couple to the electronic transition (the analogs of the mode considered above). A secondary set of thermal bath modes, assumed to be in thermal equilibrium, are taken to couple linearly to the primary modes. The dynamics of the primary modes is considered explicitly, i.e. unlike most former treatments their non-equilibrium state is derived from the dynamics of the problem. As such, they are driven by the electronic current and relax due to their coupling both to the electronic system and to the thermal bath of secondary modes. This aspect of the formalism is probably not very important for most IETS observations, but it makes it possible to address the issue of heating and subsequent possible chemical rearrangements in the junction.[41] Finally, the inelastic tunneling problem is considered in the self-consistent Born approximation[42,43] that goes far beyond the second order approximation used earlier. This makes it possible to account for overtones of the inelastic signals that are observed in resonant tunneling situations. Furthermore, we show that while the second order approximation captures much of the essential physics of the IET process, infinite order corrections can lead to quantitative differences with qualitative implications, e.g. peaks (dips) in $d^2I/d\Phi^2$ vs. $\Phi$ predicted by the low order theory can appear as dips (peaks) in the infinite order calculation. Finally, experimentally verifiable predictions are made with respect to the dependence of the *shape* of the vibrational features in $d^2I/d\Phi^2$ as the molecule-lead coupling is changed e.g. by changing the distance of the STM tip.

Our general model is described in Section 2 where an outline of the NEGF method and the self consistent Born approximation is also presented. In section 3 we give some representative results in which we (a) compare the predictions of different approximation schemes, (b) demonstrate the presence of inelastic contributions in the normal tunneling current (second derivative spectrum) and as sidebands to resonant peaks in the conductance (first derivative) spectrum, and (c) show the effect of junctions characteristics, in particular the tip-molecule distance as expressed by the corresponding electron escape rate, on the IETS lineshape. In Section 4 we discuss the linewidth of the IETS vibrational features in conjunction with the recent experimental results of Wang et al[18] and assess the relative importance of different contributions to the observed "intrinsic" linewidth. Section 5 concludes.

## 2. Technical details

We consider a two terminal junction with leads represented by free electron reservoirs in thermal equilibrium coupled through a bridging molecular system. The assumption that the electrodes are in thermal equilibrium under the steady-state operation of the junction corresponds to a weak coupling situation (i.e. conduction much smaller than $\pi e^2/\hbar$) which characterizes most molecular junctions. In what follows we refer to the molecular bridge (possibly with a few lead atoms on both sides, constituting together an extended molecule) as our system. Nuclear motions are described as harmonic normal modes and are divided into two groups. The primary group includes local phonons that interact with the electronic system. Electron-phonon interaction in the leads is disregarded. The secondary phonon group represents the environment, assumed to be in thermal equilibrium, which is coupled to the local phonons. The primary phonons are thus driven by the non-equilibrium electronic system concurrently with interchanging energy with their thermal environment. The zero-order Hamiltonian in second quantization takes the form

$$\hat{H}_0 = \sum_{i,j} t_{ij}\hat{c}_i^\dagger \hat{c}_j + \sum_{k \in L,R} \varepsilon_k \hat{d}_k^\dagger \hat{d}_k + \sum_l \Omega_l \hat{a}_l^\dagger \hat{a}_l + \sum_m \omega_m \hat{b}_m^\dagger \hat{b}_m \qquad (3)$$

with $t_{ii} = E_i$. The four terms on the right hand side represent respectively electrons on the molecules, electrons on the leads, the primary subset of molecular harmonic modes and the secondary subset of harmonic modes representing the thermal environment. The first and third terms are obvious generalizations of the corresponding terms in Eq. (1) allowing for many primary phonon modes and for many molecular electronic states. In the last term $\hat{b}_m\left(\hat{b}_m^\dagger\right)$ represent the annihilation (creation) operators of the phonon bath modes. The single electron basis chosen to represent the molecular electronic system can vary: $\hat{c}_j\,(\hat{c}_j^\dagger)$ can correspond to an atomic or molecular orbital, a lattice point, a plane wave or any other convenient basis. Also, any additional single electron term such as the effect of an external field can be incorporated into the first term of Eq. (3). Internal degrees of freedom such as spin are assumed to be incorporated in the indices used.

This zero order description is supplemented by the interaction Hamiltonian

$$\hat{H}_1 = \sum_{k \in L,R;i} \left(V_{ki}\hat{d}_k^\dagger \hat{c}_i + h.c.\right) + \sum_{l,i} M_i^l \hat{A}_l \hat{c}_i^\dagger \hat{c}_i + \sum_{l,m} U_m^l \hat{A}_l \hat{B}_m \qquad (4)$$

where $\hat{A}_l = \hat{a}_l^\dagger + \hat{a}_l$ and $\hat{B}_m = \hat{b}_m^\dagger + \hat{b}_m$. The three terms in Eq. (4) correspond respectively to coupling to the leads, coupling of the primary phonons to the electronic system (here taken to be of the polaronic form) and interaction of the local phonon modes with their thermal environment.

The principal objects used in the dynamical description of a coupled many body quantum system within the NEGF approach are the one-particle Green's functions (GFs). In our coupled electron-phonon system these are the electronic and phononic GFs on the molecular bridge defined on the Keldysh contour by

$$G_{ij}(\tau,\tau') = -i\left\langle \hat{T}_c \hat{c}_i(\tau)\hat{c}_j^\dagger(\tau')\right\rangle \quad \text{(electrons)} \tag{5}$$

$$D_{ij}(\tau,\tau') = -i\left\langle \hat{T}_c \hat{A}_i(\tau)\hat{A}_j^\dagger(\tau')\right\rangle \quad \text{(phonons)} \tag{6}$$

where $\hat{T}_c$ is the contour ordering operator. These GFs satisfy Dyson-type equations

$$G(\tau,\tau') = G_0(\tau,\tau') + \int_c d\tau_1 \int_c d\tau_2\, G_0(\tau,\tau_1)\Sigma(\tau_1,\tau_2)G(\tau_2,\tau') \tag{7}$$

$$D(\tau,\tau') = D_0(\tau,\tau') + \int_c d\tau_1 \int_c d\tau_2\, D_0(\tau,\tau_1)\Pi(\tau_1,\tau_2)D(\tau_2,\tau') \tag{8}$$

where $\Sigma$ and $\Pi$ are the electron and phonon self energies, respectively. These are essentially dressed single-particle interactions that can be represented to a desired level of approximation by the GFs themselves, thus providing a closed set of equations that can be solved self-consistently. The standard way for treating steady-states proceeds by projecting these equations onto the real time axis, yielding equations for the projected GFs and self energies (denoted as usual by the superscripts a, r, < and >), and using the fact that at steady states all two-time quantities depend only on the time difference to obtain the corresponding equations in Fourier (energy) space. The resulting equations for the GFs are the Dyson equations for the retarded GFs

$$G^r(E) = \left(\left[G_0^r(E)\right]^{-1} - \Sigma^r(E)\right)^{-1} \tag{9}$$

$$D^r(\omega) = \left(\left[D_0^r(\omega)\right]^{-1} - \Pi^r(\omega)\right)^{-1} \tag{10}$$

(and the equivalent equations for their advanced counterparts) and the Keldysh equations for the lesser and greater projections

$$G^<(E) = G^r(E)\Sigma^<(E)G^a(E) \quad \text{(and same with } <\leftrightarrow>) \tag{11}$$

$$D^<(\omega) = D^r(\omega)\Pi^<(\omega)D^a(\omega) \quad \text{(and same with } <\leftrightarrow>) \tag{12}$$

In Eqs. (9) and (10) $G_0^r(E)$ and $D_0^r(\omega)$ are the electron and local phonon Green functions for the uncoupled system described by the Hamiltonian (3). We have denoted the Fourier variables associated with the electronic and phononic GFs by $E$ and $\omega$ respectively. In the non-crossing approximation[44] (which in the present context amounts to assuming that the interactions of the 'system' with different 'bath' environments are independent of each other) the self energies are obtained in the forms

$$\Sigma(E) = \Sigma^L(E) + \Sigma^R(E) + \Sigma^{ph}(E) \tag{13}$$

$$\Pi(\omega) = \Pi^{ph}(\omega) + \Pi^{el}(\omega) \tag{14}$$

The components of the electronic self energy $\Sigma$ are those associated with the coupling to the left and right leads, $\Sigma^L$ and $\Sigma^R$, and that arising from the coupling to the primary phonons, $\Sigma^{ph}$. The former can be obtained exactly, leading to particularly simple forms in the wide band limit for the leads that essentially implies that the real parts of the corresponding retarded (advanced) self energies can be disregarded while their imaginary parts are approximated as energy independent constants:

$$\left[\Sigma_{L(R)}^r\right]_{ij} = \left[\Sigma_{L(R)}^a\right]_{ji}^* = -(1/2)i\Gamma_{ij}^{L(R)} \tag{15a}$$

$$\left[\Sigma_{L(R)}^<\right]_{ij} = i\Gamma_{ij}^{L(R)} f_{L(R)}(E) \tag{15b}$$

$$\left[\Sigma_{L(R)}^>\right]_{ij} = -i\Gamma_{ij}^{L(R)}\left[1 - f_{L(R)}(E)\right] \tag{15c}$$

where $f_{L(R)}(E)$ are the Fermi-Dirac distributions characterized by the corresponding chemical potentials $\mu_{L(R)}$, $f_K(E) = \left[\exp\left[(E - \mu_K)/k_B T\right] + 1\right]^{-1}$ and $\Gamma_{ij}^{L(R)}$ is the level-width matrix defined by

$$\Gamma_{ij}^K(E) = 2\pi \sum_{k \in K} V_{ik} V_{kj} \delta(E - E_k) \quad \text{(independent of } E \text{ in the wide-band limit)} \tag{16}$$

with $K = L, R$ denoting the left or right lead. The phonon contribution to the electronic self-energy reads

$$\left[\Sigma^r_{ph}(E)\right]_{ij} = i\sum_{k_1,k_2} M_i^{k_1} M_j^{k_2} \int \frac{d\omega}{2\pi}\Big[D^<_{k_1k_2}(\omega)G^r_{ij}(E-\omega) + D^r_{k_1k_2}(\omega)G^<_{ij}(E-\omega)$$
$$+ D^r_{k_1k_2}(\omega)G^r_{ij}(E-\omega)\Big] \quad (17a)$$
$$+ \delta_{ij}\sum_{k_1,k_2,i'} M_i^{k_1} M_{i'}^{k_2} n^{el}_{i'} D^r_{k_1k_2}(\omega=0)$$

$$\left[\Sigma^<_{ph}(E)\right]_{ij} = i\sum_{k_1,k_2} M_i^{k_1} M_j^{k_2} \int \frac{d\omega}{2\pi} D^<_{k_1k_2}(\omega)G^<_{ij}(E-\omega) \quad (17b)$$

$$\left[\Sigma^>_{ph}(E)\right]_{ij} = i\sum_{k_1,k_2} M_i^{k_1} M_j^{k_2} \int \frac{d\omega}{2\pi} D^>_{k_1k_2}(\omega)G^>_{ij}(E-\omega) \quad (17c)$$

The term containing the factor $n^{el}_{i'}$ in Eq. (17a) is the so called Hartree term in the electron-phonon interaction. Here $n^{el}_i = \rho_{ii}$ where

$$\rho_{ij} = -i\int \frac{dE}{2\pi} G^<_{ij}(E) \quad (18)$$

This Hartree term does not appear in solid-state treatments of this problem because by momentum conservation they turn out to involve only phonons of zero momentum whose density vanishes. The other terms are standard in the self-consistent Born approximation to this problem.

The self energy matrix associated with the primary phonons is expressed in Eq. (14) as a sum of contributions from the interaction with the electronic system, $\Pi^{el}(\omega)$ and the interaction with the thermal bath of secondary phonons, $\Pi^{ph}(\omega)$. The latter is obtained exactly in the wide band limit of this thermal bath, using the fact that this bath is in thermal equilibrium. Its projected components are given by [45]

$$\left[\Pi^r_{ph}(\omega)\right]_{ij} = -(1/2)i\,\mathrm{sgn}(\omega)\gamma_{ij}(\omega) \quad (19a)$$

$$\left[\Pi^<_{ph}(\omega)\right]_{ij} = -i\gamma_{ij}(\omega)F(\omega) \quad (19b)$$

$$\left[\Pi^>_{ph}(\omega)\right]_{ij} = -i\gamma_{ij}(\omega)F(-\omega) \quad (19c)$$

where

$$F(\omega) = \begin{cases} N(|\omega|) & \omega > 0 \\ 1+N(|\omega|) & \omega < 0 \end{cases} \quad (20)$$

$$\gamma_{ij}(\omega) = 2\pi \sum_m U_m^i U_m^j \delta(\omega - \omega_m) \quad (\omega \text{ independent in the wide band limit}) \quad (21)$$

and where $N(\omega) = [\exp(\omega/k_B T) - 1]^{-1}$ is the Bose Einstein distribution. Finally, the contribution $\Pi^{el}(\omega)$ to the primary phonons self-energy due to their coupling to the electronic system can be expressed in terms of the electronic Green's function. Its projections are

$$\left[\Pi_{el}^r(\omega)\right]_{ij} = -i \sum_{i_1,i_2} M_{i_1}^i M_{i_2}^j \int \frac{dE}{2\pi} \left[ G_{i_1 i_2}^<(E) G_{i_2 i_1}^a(E-\omega) + G_{i_1 i_2}^r(E) G_{i_2 i_1}^<(E-\omega) \right] \quad (22a)$$

$$\left[\Pi_{el}^<(\omega)\right]_{ij} = -i \sum_{i_1,i_2} M_{i_1}^i M_{i_2}^j \int \frac{dE}{2\pi} G_{i_1 i_2}^<(E) G_{i_2 i_1}^>(E-\omega) \quad (22b)$$

$$\left[\Pi_{el}^>(\omega)\right]_{ij} = -i \sum_{i_1,i_2} M_{i_1}^i M_{i_2}^j \int \frac{dE}{2\pi} G_{i_1 i_2}^>(E) G_{i_2 i_1}^<(E-\omega) \quad (22c)$$

The self consistent Born approximation (SCBA) is a computational scheme that effectively sums an infinite subset of non-crossing diagrams in the perturbation expansion of Green functions of many-body systems.[42,46] The procedure starts with the expressions for the Green's functions of the electronic system and the primary phonons that are zero order in the electron-phonon interaction. It is convenient to designate these as our zero order GFs for the rest of the calculation. With this re-designation Eqs. (9)-(10) remain valid provided that only $\Sigma^{ph}(E)$ and $\Pi^{el}(\omega)$ are included in the corresponding self energies. For the case of a single bridge level these zero order GFs are given by

$$G_0^r(E) = \left(G_0^a(E)\right)^* = \left[E - E_1 + (1/2)i\Gamma(E)\right]^{-1} \quad (23a)$$

$$G_0^<(E) = \frac{if_L(E)\Gamma_L(E) + if_R(E)\Gamma_R(E)}{(E - E_1)^2 + (\Gamma(E)/2)^2} \quad (23b)$$

$$G_0^>(E) = \frac{-i[1 - f_L(E)]\Gamma_L(E) - i[1 - f_R(E)]\Gamma_R(E)}{(E - E_1)^2 + (\Gamma(E)/2)^2} \quad (23c)$$

with $\Gamma = \Gamma_L + \Gamma_R$. For the case of a single primary harmonic mode the phonon GF projections are

$$D_0^r(\omega) = \left(D_0^a(\omega)\right)^* = \frac{1}{\omega - \Omega_0 + (i/2)\gamma(\omega)} - \frac{1}{\omega + \Omega_0 + (i/2)\gamma(\omega)} \quad (24a)$$

$$D_0^<(\omega) = F(\omega)\left[D_0^r(\omega) - D_0^a(\omega)\right]\text{sgn}(\omega) \tag{24b}$$

$$D_0^>(\omega) = F(-\omega)\left[D_0^r(\omega) - D_0^a(\omega)\right]\text{sgn}(\omega) \tag{24c}$$

where $F(\omega)$ is defined in Eq. (20). If (as is commonly done) the relaxation to a thermal bath of secondary phonons is disregarded, Eqs (24) take the simpler form

$$D_0^r(\omega) = \frac{1}{\omega - \Omega_0 + i\delta} - \frac{1}{\omega + \Omega_0 + i\delta} \tag{25a}$$

$$D_0^<(\omega) = -2\pi i \left\{ N(\Omega_0)\delta(\omega - \Omega_0) + (1 + N(\Omega_0))\delta(\omega + \Omega_0) \right\} \tag{25b}$$

$$D_0^>(\omega) = -2\pi i \left\{ N(\Omega_0)\delta(\omega + \Omega_0) + (1 + N(\Omega_0))\delta(\omega - \Omega_0) \right\} \tag{25c}$$

These expressions are easily generalized to situations with many electronic states and many primary phonons on the bridge. For the latter one often assumes for simplicity that they are not mixed by their interaction with the thermal bath, so Eqs. (24) are taken to hold for each mode separately.

The numerical calculation of these Green functions and self energies involves repeated integrations over the electronic energy E and the frequency variable ω. These are done using numerical grids that are chosen large enough to span the essential energy and frequency regions of the corresponding spectra, and dense enough relative to the spectral widths to yields reliable quadratures. As always, the choice of grid parameters reflect a compromise between accuracy and numerical efficiency. Some details on our choices are provided in the Figure captions in Sect. 3.

At each iteration step the Green functions of the previous step are used to update electron and phonon self-energies associated with the electron-phonon interaction using Eqs. (17) and (22), which in turn are used in (9)-(12) to obtain the next generation Green functions. The procedure terminates when the self-energies (17) and (22) have converged. Convergence of a matrix **m** is determined by the condition

$$\left| \frac{m_{ij}^{(n)} - m_{ij}^{(n-1)}}{m_{ij}^{(n-1)}} \right| < \delta \; ; \quad \text{(all } i, j\text{)} \tag{26}$$

with $\delta$ some predefined tolerance, where *n*-1 and *n* are subsequent iteration steps.

After convergence is achieved, the resulting Green functions and self-energies can be used to calculate many important one-particle characteristics of the junction. In particular, the total current through the junction is given by

$$I_{L(R)} = \frac{2e}{\hbar}\int\frac{dE}{2\pi}\text{Tr}\left[\Sigma^<_{L(R)}(E)G^>(E) - \Sigma^>_{L(R)}(E)G^<(E)\right] \qquad (27)$$

Here $I_{L(R)}$ is the current at the left (right) molecule-lead contact. It can be shown that $I_L = -I_R$ in accordance with Kirchoff's law. Using Eq. (11) and the assumed additive form (13) of the electronic self-energy, the total current (27) can be recast as a sum of elastic and inelastic contributions written below at the left contact:

$$I_{el} = \frac{2e}{\hbar}\int\frac{dE}{2\pi}\text{Tr}\Big[\Sigma^<_L(E)G^r(E)\left[\Sigma^>_L(E) + \Sigma^>_R(E)\right]G^a(E)$$
$$-\Sigma^>(E)G^r(E)\left[\Sigma^<_L(E) + \Sigma^<_R(E)\right]G^a(E)\Big] = \qquad (28)$$
$$= \frac{2e}{\hbar}\int_{-\infty}^{\infty}\frac{dE}{2\pi}(f_L(E) - f_R(E))\text{Tr}\left[\Gamma_L(E)G^r(E)\Gamma_R(E)G^a(E)\right]$$

$$I_{inel} = \frac{2e}{\hbar}\int\frac{dE}{2\pi}\text{Tr}\left[\Sigma^<_L(E)G^r(E)\Sigma^>_{ph}(E)G^a(E) - \Sigma^>_L(E)G^r(E)\Sigma^<_{ph}(E)G^a(E)\right] \qquad (29)$$

The identification of these contributions to the total current with the elastic and inelastic components can be substantiated by considering the equivalent scattering forms of these equations (see Appendix A). Note that the "Landauer form" obtained in (28) contains retarded and advanced GFs that are renormalized by the electron-phonon interaction. In fact, this renormalization yields the phonon induced correction to the elastic current. In the lowest order in the electron-phonon interaction ($2^{nd}$ order in the coupling $M$) the GFs in (29) are replaced by their zero-order counterparts (23a) and $\Sigma_{ph}$ is taken from the lowest order equivalent of Eq. (17) in which the GFs $G$ and $D$ are represented by their zero order counterparts:

$$I^{(2)}_{inel} = \frac{2e}{\hbar}\int\frac{dE}{2\pi}\text{Tr}\left[\Sigma^<_L(E)G^r_0(E)\Sigma^>_{ph,0}(E)G^a_0(E) - \Sigma^>_L(E)G^r_0(E)\Sigma^<_{ph,0}(E)G^a_0(E)\right]$$
$$(30)$$

To obtain Eq. (28) in the same order, the GFs are expressed by the lowest order Dyson forms

$$G^r(E) = G^r_0(E) + G^r_0(E)\Sigma^r_{ph,0}(E)G^r_0(E) \quad \text{(and same for } r \leftrightarrow a\text{)} \qquad (31)$$

to get

$$I_{el}^{(0)} + I_{el}^{(2)} = \frac{2e}{\hbar} \int_{-\infty}^{\infty} \frac{dE}{2\pi} (f_L(E) - f_R(E)) Tr\left[\Gamma_L(E) G_0^r(E) \Gamma_R(E) G_0^a(E)\right] +$$
$$\frac{2e}{\hbar} \int_{-\infty}^{+\infty} \frac{dE}{2\pi} (f_L(E) - f_R(E)) Tr\left[\Gamma_L(E) G_0^r(E) \Sigma_{ph,0}^r(E) G_0^r(E) \Gamma_R(E) G_0^a(E) + h.c.\right] \quad (32)$$

Eqs. (30) and (32) were recently used by Mii et al[31] to rederive the results of Persson and Baratoff[34,35] for the IETS spectra for a model of a single electronic level connecting between the leads (in this case all GFs and self energies are scalars and the trace operation in (30) and (32) is unneeded). In appendix B we repeat this derivation along a somewhat different route that yields the phonon-induced correction $\delta I$ to the current obtained in the absence of electron-phonon coupling in a form that is not limited (as the earlier work) to the lowest (second) order in the electron-phonon interaction. This improved representation will be useful in the discussion of Section 4 below of IETS linewidths. For a one-level bridge this calculation leads to

$$\delta I = \frac{2e}{\hbar} \int_{-\infty}^{+\infty} dE \frac{\Gamma_L \Gamma_R}{\Gamma^2} \rho_{el}(E) \Gamma_{ph}(E) \frac{\left(E - \tilde{E}_1(E)\right)^2 - (\Gamma/2)^2}{\left(E - \tilde{E}_1(E)\right)^2 + (\Gamma/2)^2} [f_L(E) - f_R(E)] \quad (33)$$

where $\rho_{el}(E) = -\operatorname{Im} G^r(E)/\pi$ is the electronic density of states, $\Gamma_{ph}(E) = -2\operatorname{Im}\Sigma_{ph}^r(E)$ and $\tilde{E}_1(E) = E_1 + \operatorname{Re}\Sigma_{ph}^r(E)$. Eq. (33) is similar to Eq. 17 of Ref. [31], except that $\Sigma_{ph,0}(E)$ (the lowest order phonon contribution to the electronic self-energy) and its real and imaginary components, are replaced here by by $\Sigma_{ph}(E)$ and its components, and $\rho_{el,0}(E)$, the zero order electronic density of states is replaced by its exact counterpart.

We note in passing that another important quantity readily calculated from these GFs and self energies is the power loss in the junction, i.e. the net energy flux from the electronic into the phononic system

$$P = -\frac{2}{\hbar} \int \frac{dE}{2\pi} E \operatorname{Tr}\left[\Sigma_{ph}^<(E) \mathbf{G}^>(E) - \Sigma_{ph}^>(E) \mathbf{G}^<(E)\right] \quad (34)$$

Some preliminary results based on this expression are shown below, however we defer a thorough analysis of this issue to a separate publication.

To end this section we consider the different levels of approximations that can be used for describing IETS. The procedure described above constitutes the self-consistent

Born approximation. In the simple Born approximation one replaces all the GFs in Eqs. (28) and (29) by their expression (9)-(12), however the self energies $\Sigma_{ph}$ and $\Pi_{el}$ are calculated from Eqs. (17) and (22) using the zero order GFs. This amounts to stopping the SCBA procedure after the first iteration. Finally, Eqs. (30) and (32) represent results obtained in the lowest order in the electron-phonon interaction. In what follows we refer to these approximations as the Self consistent Born Approximation (SCBA), the Simple Born Approximation (BA) and the lowest order perturbation theory (LOPT).

## 3. Representative numerical results

In the calculations described below we have employed a model with a single harmonic mode in the primary set directly coupled to the electronic system. We primarily consider two situations that are distinguished by the choice of molecule-electrodes couplings. In one that corresponds to strong chemisorption to both leads we take $\Gamma_L = \Gamma_R = 0.5$eV. In the other, more akin to an STM geometry, we take $\Gamma_L$=0.05eV $\Gamma_R$=0.5eV, where the more weakly coupled electrode represents physisorption. The Fermi energies of the two leads were taken to be zero in the unbiased junction. As a "standard" set of parameters we have also taken the electronic state energy in the unbiased junction $E_i=t_{ii}$=1eV, the primary phonon frequency $\Omega_0 = 0.13$ eV, the electron-phonon coupling $M$=0.3eV and the thermal relaxation rate of the phonon $\gamma_{ph} = 1 \cdot 10^{-3}$ eV. Under bias $\Phi$ we kept the energy $E_1$ of a single site bridge fixed and took

$$\begin{aligned} \mu_L &= E_F + \eta |e| \Phi \\ \mu_R &= E_F - (1-\eta)|e| \Phi \end{aligned} \quad (35)$$

where $e$ is the electron charge and $E_F$ is the unbiased Fermi energy that is set to zero. $\eta$ is the "voltage division factor"[47] chosen as $\eta = \Gamma_R / \Gamma$ with $\Gamma = \Gamma_L + \Gamma_R$. This reflects a choice of model in which the molecular state is essentially pinned to the Fermi level of the chemically bonded electrode in the STM configuration where $\Gamma_L \ll \Gamma_R$. When a model of two bridge electronic states was used, the corresponding atomic energies $E_i = t_{ii}$ were taken to interpolate linearly between $\mu_L$ and $\mu_R$, and their interstate coupling was chosen $t_{12} = 0.3$ eV. Note that the choice $M$=0.3eV for the electron-phonon coupling corresponds to a reorganization energy of $M^2/\Omega_0 \approx 0.7$ eV. These parameters were selectively varied in different calculations in order to elucidate their effect on the inelastic signal. Finally, the convergence tolerance for self-consistent procedure (Eq.(18)) was set to $10^{-4}$.

Results for the current-voltage behavior of junctions characterized by these parameters at temperature $T$=300K are shown in Figures 2a and 3a. Fig. 2 depicts results obtained for the single level bridge while Fig. 3 shows similar results obtained for the 2-level bridge. These results compare the completely elastic case (without coupling to phonons) with the results obtained in the presence of electron-phonon coupling calculated both in the limit where thermal relaxation of the primary phonon is fast (so that it may be taken to be in thermal equilibrium and the thermal (secondary) phonon bath may be disregarded), and in the more realistic case when this phonon reaches a steady state balanced between pumping by the electronic system and relaxing into its thermal environment. We note in passing that the appearance of negative differential resistance in the ballistic transport shown in Fig. 3a results from our model assumption of a linear potential drop on the bridge that, for large bias, has the effect of shifting the two bridge levels out of resonance with each other. We see that electron-phonon coupling makes only a modest effect on the current-voltage behavior even for the present choice of relatively strong coupling. However the main interest in these plots lies in the power loss that results in heating of the phonon system. Figs. 2b,c and 3b,c show the absolute and the relative (with respect to the overall bias) power loss, Eq. (34), and the oscillator "temperature" (defined by equating its average steady state occupation to the corresponding Bose Einstein expression) obtained for the chosen parameters from the full self-consistent calculation as functions of the applied voltage. The relatively high steady state temperature obtained ($T$=4000K corresponds to an average phonon occupation $n$~2) reflects the relatively large coupling used and the fact that the full extent of the electron phonon-coupling (as expressed by the choice of reorganization energy) was implemented into a single oscillator. A systematic investigation of the application of this model to the issue of junction heating will be published separately.

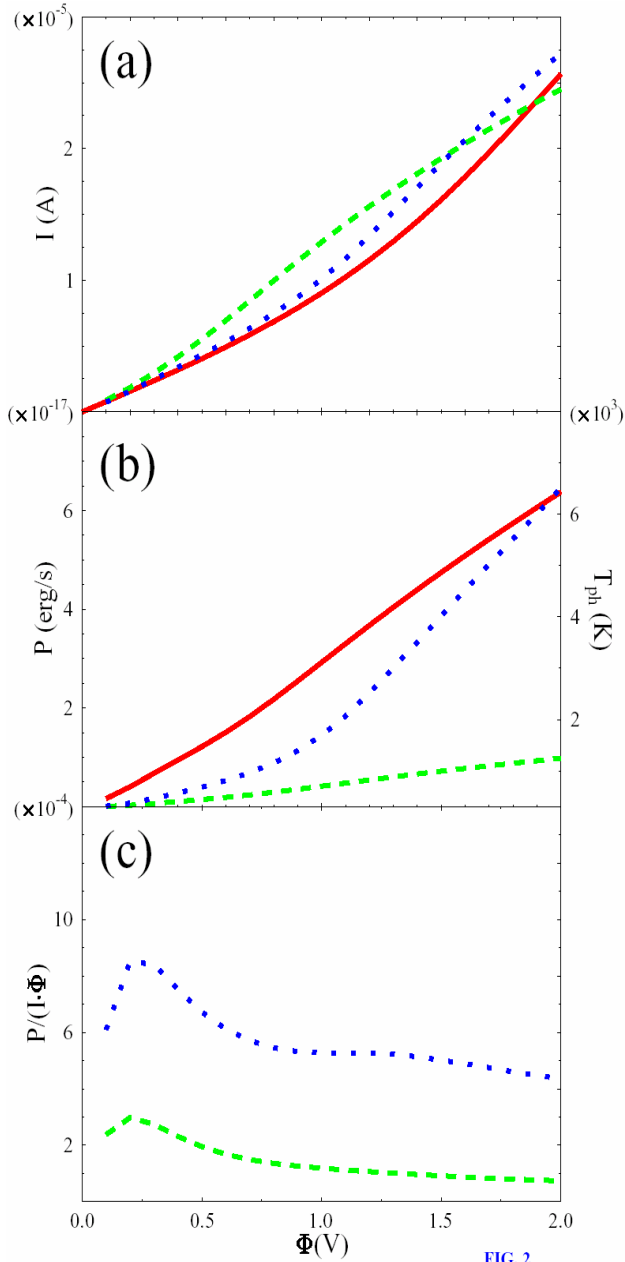

Fig. 2. (a) The current-voltage dependence in a junction characterized a single bridge level and by the "standard" set of parameters (see text) with $\Gamma_L = \Gamma_R = 0.5$eV. Full line (red) – result obtained for *M*=0 (no electron-phonon coupling). Dashed and dotted lines show the effect of electron-phonon coupling, where the dotted (blue) line represents the result obtained under the assumption that the molecular vibration is at thermal equilibrium unaffected by its coupling to the non-equilibrium electronic system while the dashed (green) line corresponds to the case where relaxation rate of the molecular vibration due to its coupling to its thermal environment (secondary phonons) is finite. Integrations over the electronic energy axis were done using an energy grid of 3501 points spanning the region between –0.5 and 3 eV with step-size 0.001 eV and those over the phonon frequency were carried using a grid of 601 points spanning the frequency range between -0.3 and 0.3 eV with step-size 0.001 eV. (b) The power loss (dashed (green) line, left axis) and the non-equilibrium oscillator 'temperature' (full (red) line, right axis) for the case represented by dashed line in (a) plotted against the voltage bias. The dotted (blue) line depicts the power loss for the case represented by the dotted (blue) line in (a). (c) Same as the dashed line in (b), where the power loss is displayed relative to the total available power *I*Φ.

FIG. 2

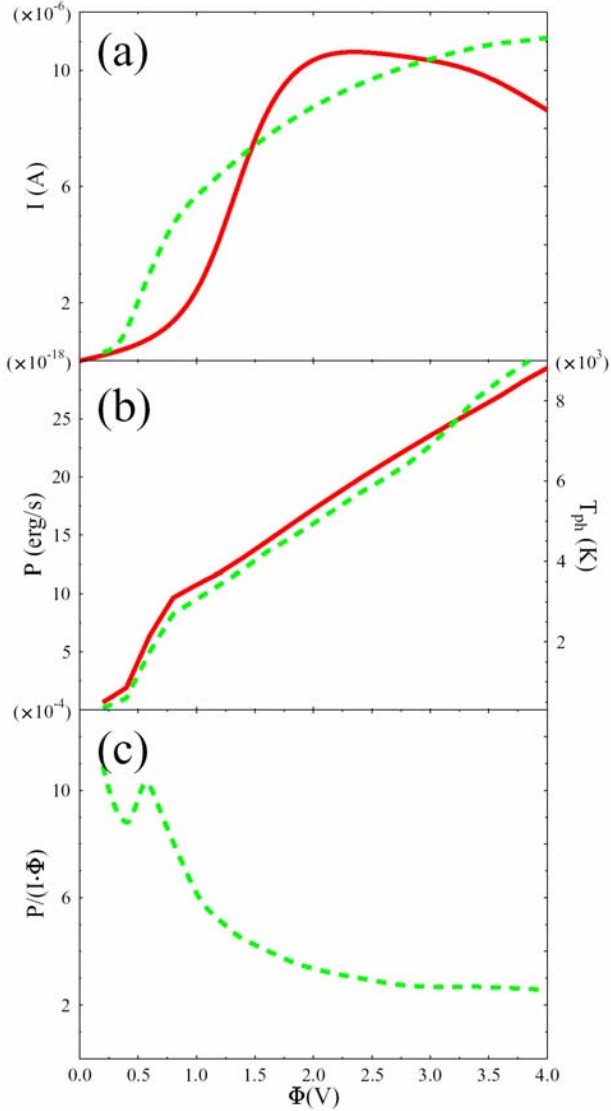

Fig. 3. Same as Fig. 2, for a two electronic state bridge (see text for the 'standard' parameters), comparing the ballistic case and the case with electron-phonon coupling and finite thermal relaxation rate.

Figure 4 repeats the calculation of Fig. 2 at temperature $T$=10K, now focusing on the inelastic signal. Here we show the calculated inelastic tunneling spectrum, $d^2I/d\Phi^2$ plotted against the bias potential $\Phi$, and compare the results obtained from the three computational schemes discussed in Section 2. It is seen that the signal obtained from the LOPT approach as well as from the BA calculation is underestimated at low voltage and overestimated at the larger voltage regime relative to the full SCBA calculation. Furthermore, while the three calculations describe rather well the first phonon peak, higher harmonics are seen most strongly in the full self-consistent (SCBA) treatment and faintly in the BA calculation.

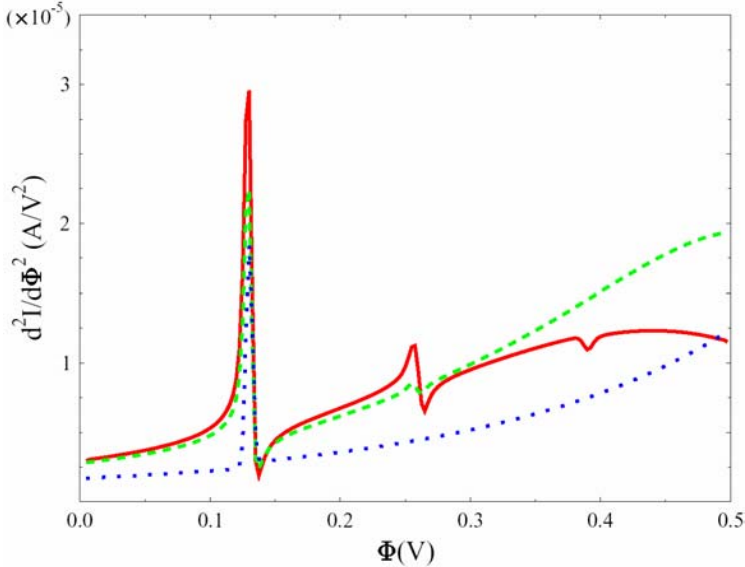

Fig. 4. The inelastic tunneling spectrum, $d^2I/d\Phi^2$ plotted against the bias potential $\Phi$, for the model with a single bridge state at $T$=10K. The 'standard' parameters are used with $\Gamma_L$=0.05eV $\Gamma_R$=0.5eV. The calculation is done using the SCBA (full line, red), the BA (dashed line, green) and the LOPT (dotted line, blue) schemes. These calculations were done using a grid in electron energy of 1501 points spanning range from -1.0 eV to 2.75 eV with steps size 0.0025 eV and a grid in the phonon frequency of 201 points spanning range from –0.5 eV to 0.5 eV with step 0.0025 eV.

As discussed in Section 1, a different kind of inelastic signal is obtained as phonon side bands to the resonant tunneling peaks in the conductance/voltage plot. For bridge assisted tunneling such peaks are expected at the bias potential for which the molecular HOMO or LUMO level just enters the potential window between $\mu_L$ and $\mu_R$ (and similarly at Coulomb blockade thresholds). Phonon sidebands should appear as additional peaks, that may be resolved if the resonance width, of order $\Gamma = \Gamma_L + \Gamma_R$, is small relative to the phonon frequency $\Omega_0$. Figure 5 shows this effect, where the same set of standard parameters was used as in Fig. 2 except that we take $\Gamma_L = \Gamma_R$ =0.013eV. For this choice of parameters, however, the BA calculation failed while the SCBA procedure did not converge.[48] The results shown in Fig. 5 are obtained using the scattering theory approach of Wingreen et al,[51] where a small polaron transformation is applied to affect a high order treatment of the electron-phonon coupling. This calculation yields the transmission coefficient $T(E)$ as a function of the electron energy. At low temperature and for an STM configuration with the tip acting as the left electrode and the bridge levels pinned to the substrate (right electrode), $T(E_{FL}+e\Phi)$ is directly related to the differential conduction $dI/d\Phi$. It is seen that the resonance peak associated with the bridge electronic level is accompanied by higher phonon harmonics in the transmission probability, in correspondence with experimental observations.[17]

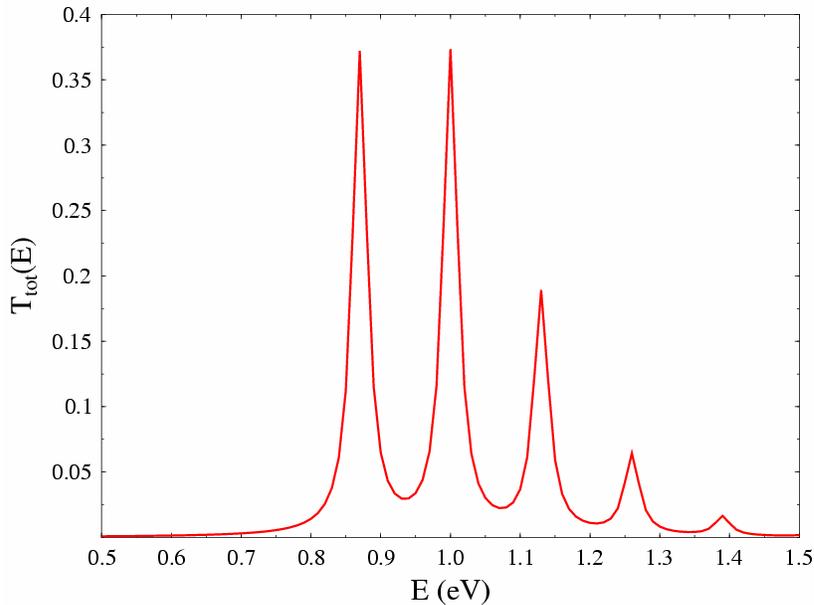

Fig. 5. The transmission coefficient T plotted as a function of electron energy $E$ for the junction of Fig. 2 where the electron escape rates $\Gamma_L$ and $\Gamma_R$ are taken each to be 0.013eV. Integration grids are the same as in figs. 2 and 3.

Next we consider situations where dips in the inelastic tunneling spectrum may be observed. This issue was discussed extensively by Persson and Baratoff[33-35] and by Ueba and coworkers[31,32] however some new observations can be made. Figure 6 compares, for the single resonant level model, results obtained using the LOPT, BA and SCBA calculation schemes. Here we use $\Omega_0$=0.13eV, $E_1$=0.6eV and $M$=0.3eV for the parameters that characterize the bridge, $T$=10K, and $\Gamma_L$=0.05eV $\Gamma_R$=0.5eV for the parameters associated with the bridge-lead coupling, and assume that the molecular vibration maintains its thermal equilibrium in the biased, current carrying junction. We indeed see a dip in both the SCBA and BA results for the inelastic tunneling spectrum at $e\Phi = \hbar\Omega_0$, however the LOPT used in the earlier works to analyze this phenomenon yields a peak rather than a dip in this spectrum. This observation is hardly surprising: The phonon-induced correction to the tunneling current includes elastic and inelastic contributions, where the former may be negative. The total effect on the tunneling spectrum depends on the balance between these positive and negative corrections, leading to peaks or dips in the second derivative spectrum. Thus, a quantitative error in the low order calculation may translate into a qualitative difference, predicting a peak rather than a dip (or vice verse) in the tunneling spectrum. The different behavior predicted by the BA and the SCBA for the higher harmonic features stems from a similar reason.

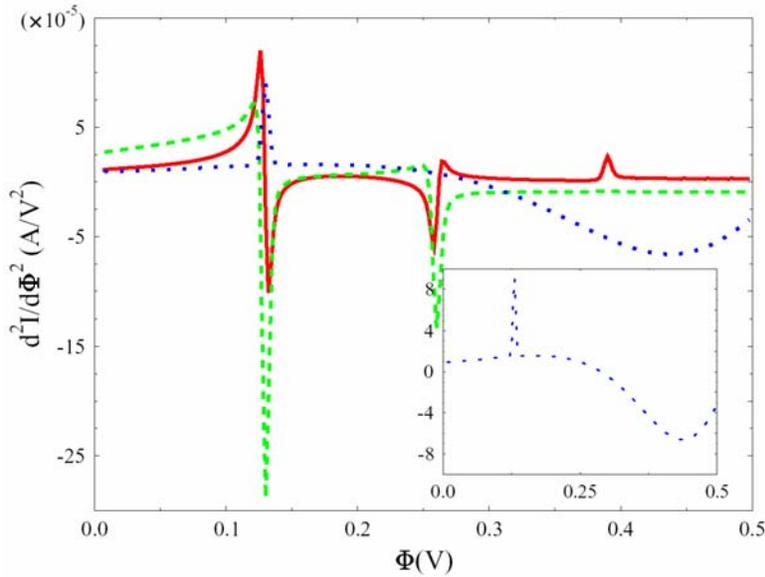

Fig. 6. $d^2I/d\Phi^2$ plotted against $\Phi$ for the single resonant level model with parameters described in the text. Full line (red) – SCBA. Dashed line (green): BA Dotted line (blue): LOPT. The inset shows an expanded view of the LOPT result. The integration grids were 3001 points from -1.0eV to 2.0eV with step size 0.001 eV for the electronic energy and 1001 points from -0.5eV to 0.5eV with step size 0.001 eV for the vibrational frequency.

The sensitivity of the observed IETS features to parameters of the molecular electronic structure and of the electron-phonon coupling was already emphasized by Mii et al.[32] An example based on the full SCBA calculation is shown in Fig. 7. Here we use the "standard" parameters with $\Gamma_L=\Gamma_R=0.5$eV and $T=10$K, and compare results for different choices of $E_1$, the position of the resonant level relative to the unbiased Fermi energies. It is seen that the character of the IETS feature can change from dip to peak as $E_1$ is increased at constant $\Gamma$, as already discussed within the LOPT calculation by Persson and Baratoff.

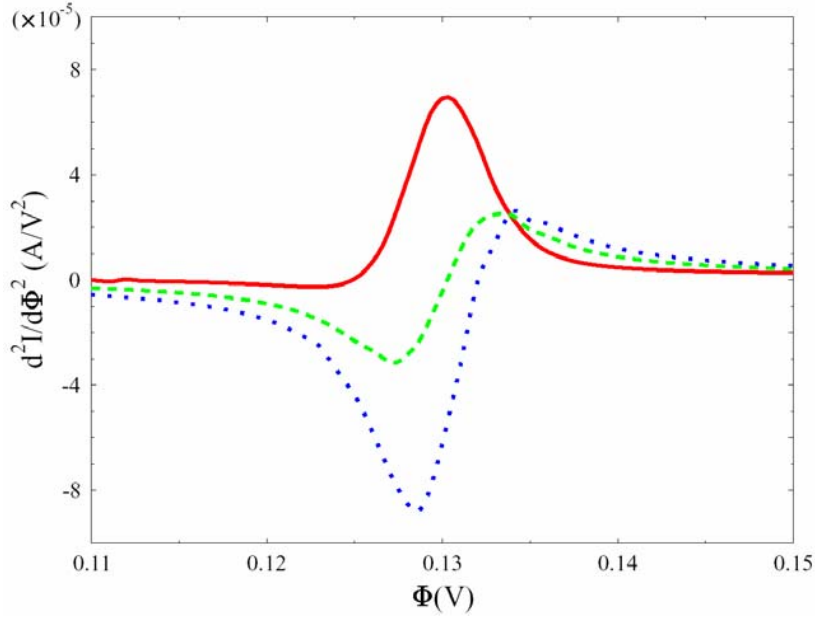

Fig. 7. A SCBA calculation of the IETS threshold feature in $d^2I/d\Phi^2$ for the one resonant level model using the 'standard' parameters (see text) with $\Gamma_L=\Gamma_R=0.5$eV and $T$=10K. Full line (red) $E_1$=0.70eV, dashed line (green) – $E_1$=0.60eV, dotted line (blue) – $E_1$=0.55eV. The grids used in the numerical calculation are the same as in Fig. 6.

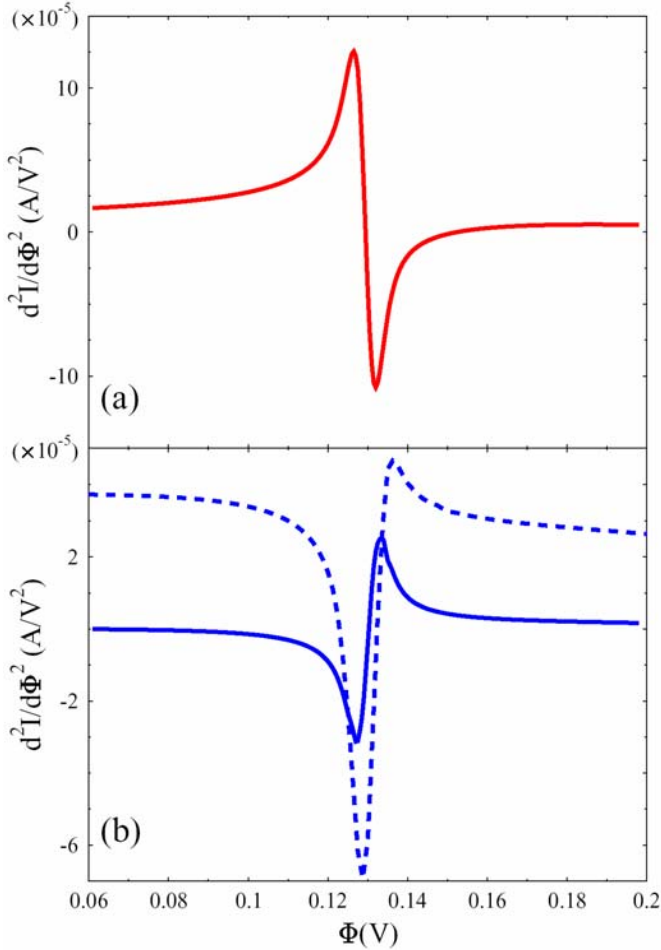

Fig. 8. A SCBA calculation of the IETS threshold feature in $d^2I/d\Phi^2$ for the one resonant level model at $T=10K$, using the 'standard' parameters (see text) except that $E_1$ is taken as 0.6eV. $\Gamma_R=0.5$eV and $\Gamma_L$ is 0.05eV in (a) and 0.5eV in (b). In (b) the full line corresponds to the case where the Fermi energies are shifted under the bias according to $\mu_L = |e|\Phi(\Gamma_R/\Gamma)$, $\mu_R = -|e|\Phi(\Gamma_L/\Gamma)$, while the dashed line was produced for the model $\mu_L = |e|\Phi$, $\mu_R = 0$. Numerical grids are as in Fig. 6.

The same observation may be carried further, into an experimentally verifiable prediction. In figure 8 we consider the same one-level molecular model in an STM configuration, so that $\Gamma_L$, taken to be the width parameter associated with the molecule-tip coupling, can be varied by changing the tip-molecule distance. Fig. 8 shows the dependence of the calculated spectrum on $\Gamma_L$ and on the potential distribution in the junction. A substantial effect on the shape of the IETS feature is seen, changing here from an essentially symmetric dip in one case to a peak-derivative like feature in the other. Indeed, both symmetric[16] and asymmetric[18] IETS dips were observed in molecular junctions and it would be of interest to systematically examine the lineshape evolution as a function of tip-substrate distance.

## 4. The IETS linewidth

Spectroscopic linewidths are often difficult to interpret since their origin may lie in diverse physical factors. In a recent IETS experiment by Wang et al[18] it was found possible to eliminate or to estimate some of the important contributions of the thermal Fermi distribution in the substrate and of the distribution of the local electrostatic field, and to come up with what the authors call an intrinsic linewidth of 3.73±0.98meV at T=4K in a junction containing a layer of alkanethiol molecules. The rate of vibrational relaxation due to nuclear coupling with the thermal environment is expected not to exceed a few wavenumbers, say up to 1meV. Inhomogeneous broadening is always a possible contribution to the observed linewidth. However for a molecule adsorbed on a metallic substrate another channel of relaxation involves the vibronic coupling to the continuum of electron hole pairs in the metal. Indeed, such coupling has been shown to be important and sometimes dominating source of broadening in the infrared spectra of molecules adsorbed on metal surfaces.[34] In this Section we use the theoretical tools presented in Section 2 to estimate the effect of vibronic coupling on the linewidth of vibrational features in IETS. Since a truly intrinsic linewidth can be observed only in a single molecule measurement, the relevant energy scales are those suitable to an STM experiment, i.e. $\Gamma_L \ll \Gamma_R$ and $E_1$ pinned to the Fermi energy of the right electrode. However in what follows we consider the general case represented by keeping $E_1$ pinned to the unbiased Fermi energy and moving the chemical potentials of the left and right electrodes according to Eq. (35). As in Sect. 3 the model $\eta = \Gamma_R/\Gamma$ was applied. The STM limit, with the left electrode representing the tip, is given by $\eta \to 1$.

We start with the $T=0$ limit of Eq. (33). It contains an explicit dependence on the imposed potential that leads to[31]

$$\frac{d(\delta I)}{d\Phi} = \frac{2e^2}{\hbar} \frac{\Gamma_L \Gamma_R}{\Gamma^2} \left\{ \eta \rho_{el}(\mu_L) \Gamma_{ph}(\mu_L) \frac{\left(\mu_L - \tilde{E}_1(\mu_L)\right)^2 - (\Gamma/2)^2}{\left(\mu_L - \tilde{E}_1(\mu_L)\right)^2 + (\Gamma/2)^2} \right.$$
$$\left. + (1-\eta) \rho_{el}(\mu_R) \Gamma_{ph}(\mu_R) \frac{\left(\mu_R - \tilde{E}_1(\mu_R)\right)^2 - (\Gamma/2)^2}{\left(\mu_R - \tilde{E}_1(\mu_R)\right)^2 + (\Gamma/2)^2} \right\} \quad (36)$$

It is seen that in (36) the dependence on the potential enters in several places, however the potential dependence of the terms $\Gamma_{ph}(E = \mu_L(\Phi))$ and $\Gamma_{ph}(E = \mu_R(\Phi))$, associated with the contribution of the electron-phonon interaction to the electronic self energy, is of special importance because of its singular character. It may be shown,[31] starting from Eq. (B2) and using (17b) and (17c) that these factors are given by[31] (see also Appendix B)

$$\Gamma_{ph}(\mu_L) = 2\pi M^2 \int_0^{|e|\Phi} d\omega \rho_{ph}(\omega) \frac{\Gamma_R}{\Gamma} \rho_{el}(\mu_L - \omega)$$
$$\Gamma_{ph}(\mu_R) = 2\pi M^2 \int_0^{|e|\Phi} d\omega \rho_{ph}(\omega) \frac{\Gamma_L}{\Gamma} \rho_{el}(\mu_R + \omega) \quad (37)$$

where $\rho_{el}(E)$ and $\rho_{ph}(\omega)$ are the electron and phonon density of states, which, for the one resonant-level model are given explicitly by (cf. Eqs. (B3) and (B4))

$$\rho_{el}(E) = -\frac{1}{\pi} \operatorname{Im} G^r(E) = \frac{1}{2\pi} \frac{\Gamma_{tot}(E)}{\left(E - \tilde{E}_1(E)\right)^2 + (\Gamma_{tot}(E)/2)^2} = \frac{\Gamma_{tot}(E)}{2\pi} G^r(E) G^a(E) \quad (38)$$

$$\rho_{ph}(\omega) = -\frac{\operatorname{sgn}(\omega)}{\pi} \operatorname{Im} D^r(\omega) = \frac{1}{2\pi} \frac{\gamma_{tot}(\omega)}{(\omega - \Omega_0)^2 + (\gamma_{tot}(\omega)/2)^2} \quad (39)$$

where $\tilde{E}_1(E) = E_1 + \operatorname{Re} \Sigma^r_{ph}(E)$, while

$$\Gamma_{tot}(E) \equiv \Gamma - 2 \operatorname{Im} \Sigma^r_{ph}(E) = \Gamma + \Gamma_{ph}(E) \quad (40a)$$

and

$$\gamma_{tot}(\omega) \equiv \gamma_{ph} - 2 \operatorname{Im} \Pi^r_{el}(\omega) = \gamma_{ph} + \gamma_{el}(\omega) \quad (40b)$$

are the total electron and phonon widths. In typical situations, where the molecule is strongly interacting with at least one electrode, $\Gamma_{tot} \gg \gamma_{tot}$, so $\rho_{ph}(\omega)$ represents a relatively sharp feature. Consider now the second derivative $d^2I/d\Phi^2$. The principal contribution to structure in this function of $\Phi$ comes from the derivative of $\Gamma_{ph}(E_F+|e|\Phi)$ in (36), while the other terms in (36) as well as the phonon independent term $I^{(0)}$ only contribute some background (and also determine the sign of the computed feature). Thus, using (36) and (37) we find, approximately

$$\frac{d^2\delta I}{d\Phi^2} = \frac{4\pi e^3}{\hbar}\frac{\Gamma_L\Gamma_R}{\Gamma^2}M^2\rho_{ph}(|e|\Phi)\left\{\eta\frac{\Gamma_R}{\Gamma}\rho_{el}(\mu_L)\rho_{el}(\mu_L-|e|\Phi)\frac{\left(\mu_L-\tilde{E}_1(\mu_L)\right)^2-(\Gamma/2)^2}{\left(\mu_L-\tilde{E}_1(\mu_L)\right)^2+(\Gamma/2)^2}\right.$$

$$\left.+(1-\eta)\frac{\Gamma_L}{\Gamma}\rho_{el}(\mu_R)\rho_{el}(\mu_R+|e|\Phi)\frac{\left(\mu_R-\tilde{E}_1(\mu_R)\right)^2-(\Gamma/2)^2}{\left(\mu_R-\tilde{E}_1(\mu_R)\right)^2+(\Gamma/2)^2}\right\}$$

$$+ \text{ background terms}$$

(41a)

or, for $\eta=1$ and/or $\Gamma_L\ll\Gamma_R$,

$$\frac{d^2\delta I}{d\Phi^2} = \frac{4\pi e^3}{\hbar}\frac{\Gamma_L\Gamma_R}{\Gamma^2}M^2\rho_{el}(E_F+|e|\Phi)\rho_{el}(E_F)\frac{\left(E_F+|e|\Phi-\tilde{E}_1(E_F+|e|\Phi)\right)^2-(\Gamma/2)^2}{\left(E_F+|e|\Phi-\tilde{E}_1(E_F+|e|\Phi)\right)^2+(\Gamma/2)^2}\rho_{ph}(|e|\Phi)$$

$$+ \text{ background terms}$$

(41b)

where the dependence on $\Phi$ has now been written explicitly. As could be expected, we find that the lineshape of the IETS feature is determined by the function $\rho_{ph}(\Phi)$ which peaks about $|e\Phi|=\Omega_0$ with an intrinsic width given by (40b). When $\Omega_0$ exceeds the Debye frequency of the thermal environment, the vibrational contribution $\gamma_{ph}$ to this width is strongly temperature dependent and its low temperature magnitude is expected to be far below 1mV.[52] The electronic contribution, $\gamma_{el}=-2\,\text{Im}\,\Pi_{el}(\omega)$ can be estimated from the second order expression for the electronic contribution to the phonon self energy, Eq. (22), using Eqs. (23). The result is

$$\gamma_{el}(\omega) = 2M^2\int_{-\infty}^{+\infty}\frac{dE}{2\pi}\frac{f_L(E)\Gamma_L+f_R(E)\Gamma_R}{(E-\varepsilon_1)^2+(\Gamma/2)^2}\text{Im}\frac{\Gamma-2i\omega}{(E+\omega-E_1+i\Gamma/2)(E-\omega-E_1-i\Gamma/2)} \quad (42)$$

Fig. 9 shows the results obtained using both this lowest order estimate of $\Pi$ and the value obtained form the converged SCBA calculation, plotted against the applied voltage bias.

The parameters used in this calculation, $T$=10K, $E_1$=1eV, $\Gamma_L$=$\Gamma_R$=0.5eV, $\Omega_0$=0.13eV, $M$=0.4eV and $\gamma_{ph}$=0.0001eV represent reasonable orders of magnitudes given available experimental data. At $|e|\Phi = \Omega_0 = 0.13\text{eV}$ we find this contribution to be a few mV, of the order observed by Wang et al.[18] While the result depends on the choice of parameters, this dependence is relatively mild so the result can be viewed as a reasonable order of magnitude estimate. We may conclude that broadening of the vibrational features due to coupling between molecular vibrations and substrate conduction electrons constitutes a substantial contribution to the intrinsic width observed in the experiment of Wang et al,[18] with inhomogeneous broadening possibly making a contribution of similar order of magnitude.

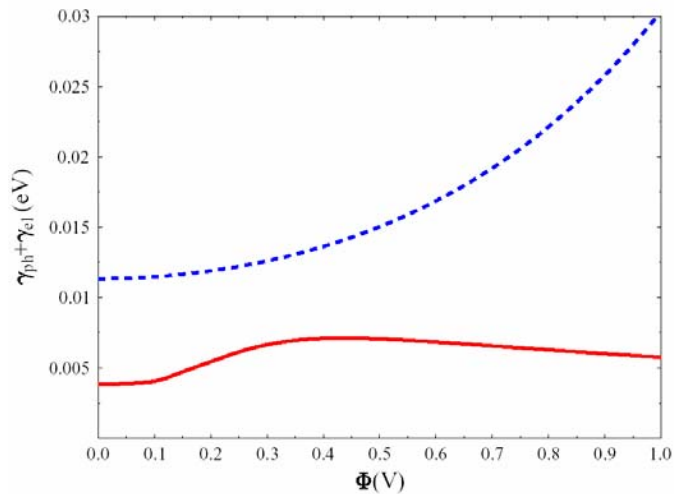

Fig. 9. The width of a vibrational IETS feature obtained from Eq. (42) (with the additional additive term $\gamma_{ph} = 0.0001\text{eV}$) plotted against the imposed potential bias. See text for parameters. Full (red) line – SCBA calculation. Dotted (blue) line – lowest order perturbation theory.

## 5. Conclusions

We have considered a scheme for phonon-assisted electron tunneling, where the mutual influence of electron and local phonon modes is taken into account and where the local phonon mode is coupled to a harmonic thermal bath. The NEGF methodology can be readily used to extend previous studies to situations involving relatively strong electron-phonon coupling. A simple resonant-level model was studied numerically, and the influence of different interactions on the junction properties was considered. We have compared results obtained using three levels of approximation: The full self consistent calculation, the Born approximation that amounts to the first step of the full procedure and the widely used perturbation theory scheme. In view of this comparison we may conclude that results based on perturbation theory have only qualitative values, and in fact their quantitative failure may translate into a qualitative one when interference phenomena such as those giving rise to dips in the tunneling spectrum play an important role in the process.

We have also noted that while the peaks and dips mentioned above are observed in second derivative of the current vs. voltage, another type of peaks - satellites of the elastic resonant peak due to phonon-assisted transport - can be observed in the first derivative (conductance) plot.

The model considered in this work assumes that under steady state conditions the electronic manifolds that describe the metal electrodes are in thermal equilibrium. This assumption is valid in the limit where the junction affects a relatively small coupling between the leads, i.e. when the conduction is much smaller than the quantum unit $e^2/\pi\hbar$. This is indeed the case for most molecular junctions. In the opposite limit encountered in metallic point contacts and also apparently in the molecular $H_2$ junction studied in Ref. [19], the transmission coefficient is close to unity, implying that the leads are not in equilibrium, e.g. backscattered electrons are locally absent in the source lead. The implication of this situation on the observed inelastic spectrum and a comparison between these two extreme cases will be considered elsewhere.

The same framework that incorporates electron-phonon interactions in the calculation of electron tunneling phenomena may be used to assess power loss and heat production during channel conductance. We have described one example of such calculation, however we defer a full discussion of this important issue to a later study.

This model and theoretical framework were also used to discuss the shape and width of vibrational features in inelastic tunneling spectroscopy. The observed lineshapes reflect contributions of phonon-induced elastic and inelastic tunneling fluxes, and are affected by interference with the phononless elastic component. This result in strong sensitivity to the energy of the resonant level, and perhaps more interestingly, to the strength of molecule-leads electronic interaction that can be controlled by the tip-molecule distance in an STM type experiment. With respect to the widths of IETS features, we have concluded that coupling of molecular vibrational motion to the conduction electrons of the lead to which the molecule binds strongly contributes a substantial part of the experimentally observed "intrinsic" linewidth of a few meV, although additional contribution from inhomogeneous broadening cannot be ruled out.

**Appendix A**

Here we give a scattering theory-like illustration that identifies the contributions (28) and (29) to the total current as the corresponding elastic and inelastic channels. In a scattering process only one state $|0\rangle$ (the incoming state) of energy $\varepsilon_0$, say in the left continuum, is assumed occupied, so the initial energy is specified. All other states $\{|l\rangle\}$ and $\{|r\rangle\}$ in the left and right continua are taken to be unoccupied so they do not contribute to the flux. Also, since we are interested in transmission rather than reflection, outscattering to the left continuum $\{|l\rangle\}$ is disregarded. Under these conditions the transport process essentially represents a decay of the state $|0\rangle$ to the (unoccupied) continuum $\{|r\rangle\}$. This implies

$$\begin{aligned}
\Sigma_0^<(E) &= 2\pi i |V_{10}|^2 \delta(E-\varepsilon_0); & \Sigma_0^>(E) &= 0 \\
\Sigma_L^<(E) &= 0 & ; & \Sigma_L^>(E) &= 0 \\
\Sigma_r^<(E) &= 0 & ; & \Sigma_r^>(E) &= -i2\pi |V_{1r}|^2 \delta(E-\varepsilon_r)
\end{aligned} \quad (A1)$$

The contribution, Eq. **(28)**, to the scattering flux from state $|0\rangle$ to state $|r\rangle$ is then

$$I_{0\to r}^{el} = \frac{2e}{\hbar} \int \frac{dE}{2\pi} \Sigma_0^<(E) G^r(E) \Sigma_r^>(E) G^a(E) = \frac{2e}{\hbar} 2\pi |V_{10}|^2 |V_{1r}|^2 G^r(\varepsilon_0) G^a(\varepsilon_0) \delta(\varepsilon_0 - \varepsilon_r)$$
(A2)

and is seen indeed to be an elastic current. Note that the GFs in (A2) are renormalized by the electron-phonon interaction. For the other contribution, Eq. (29), we get

$$I_{0 \to r}^{inel} = \frac{2e}{\hbar} \int \frac{dE}{2\pi} \Sigma_0^< (E) G^r (E) \Sigma_{ph}^> (E) G^a (E)$$

$$= -\frac{2e}{\hbar} |V_{10}|^2 M^2 G^r (\varepsilon_0) G^a (\varepsilon_0) \int_{-\infty}^{+\infty} \frac{d\omega}{2\pi} D^> (\omega) G^> (\varepsilon_0 - \omega) \quad \text{(A3)}$$

where we have used Eq. (17b). From Eqs. (23)-(25) $D^>$ is strongly peaked about $\omega = \pm \Omega_0$ so the integral is proportional to $G^> (\varepsilon_0 \pm \Omega_0)$ which is peaked about $\varepsilon_r = \varepsilon_0 \pm \Omega_0$ (to lowest order $G_0^> (E) = -i2\pi |V_{1r}|^2 \delta(E - \varepsilon_r) \left[ (E - E_1)^2 + (\Gamma(E)/2)^2 \right]^{-1}$). We see that this is indeed an inelastic contribution.

**Appendix B**

Here derive analytical approximations for the inelastic electron tunneling features that are used in the discussions of Sections 2 and 4. The expressions derived here are similar to those of Ref. [31], however they are not restricted to second-order in electron-phonon interaction, and include also phonon coupling to a thermal bath of secondary phonons.

We start from the expressions for the elastic and inelastic currents Eqs. (28) and (29). Consider first the inelastic component. For the case where $\Gamma_L(E)/\Gamma_R(E) = $ constant (which always holds in the wide band approximation) the inelastic current can be expressed in a "Landauer-like" form by taking[53] $I_{inel} = xI_{inel}^L - (1-x)I_{inel}^R$ and choosing $x = \Gamma_R / \Gamma$ where $\Gamma = \Gamma_L + \Gamma_R$. This leads to

$$I_{inel} = \frac{2e}{\hbar} \int_{-\infty}^{+\infty} \frac{dE}{2\pi} Tr \left[ \frac{\Gamma_L(E) \Gamma_R(E)}{\Gamma(E)} G^r(E) \Gamma_{ph}(E) G^a(E) \right] (f_L(E) - f_R(E)) \quad \text{(B1)}$$

with

$$\Gamma_{ph}(E) = -2 \operatorname{Im} \Sigma_{ph}^r (E) = i \left[ \Sigma_{ph}^> (E) - \Sigma_{ph}^< (E) \right] \quad \text{(B2)}$$

Henceforth, we focus on the case of a one electronic level bridge. In this case electron and phonon densities of states take the form

$$\rho_{el}(E) = -\frac{1}{\pi}\operatorname{Im} G^r(E) = \frac{1}{2\pi}\frac{\Gamma_{tot}(E)}{\left(E-\tilde{E}_1(E)\right)^2 + \left(\Gamma_{tot}(E)/2\right)^2} = \frac{\Gamma_{tot}(E)}{2\pi}G^r(E)G^a(E) \quad \text{(B3)}$$

$$\rho_{ph}(\omega) = -\frac{\operatorname{sgn}(\omega)}{\pi}\operatorname{Im} D^r(\omega) = \frac{1}{2\pi}\frac{\gamma_{tot}(\omega)}{\left(\omega-\Omega_0\right)^2 + \left(\gamma_{tot}(\omega)/2\right)^2} \quad \text{(B4)}$$

where $\tilde{E}_1(E) = E_1 + \operatorname{Re}\Sigma^r_{ph}(E)$, $\Gamma_{tot}(E) \equiv \Gamma - 2\operatorname{Im}\Sigma^r_{ph}(E) = \Gamma + \Gamma_{ph}(E)$ and $\gamma_{tot}(\omega) \equiv \gamma - 2\operatorname{Im}\Pi^r_{el}(\omega) = \gamma + \gamma_{el}(\omega)$ are total electron and phonon widths, respectively. Next we define the non-equilibrium electron and phonon occupations $n(E)$ and $N(\omega)$ according to

$$G^<(E) = 2\pi i n(E)\rho_{el}(E); \quad G^>(E) = -2\pi i[1-n(E)]\rho_{el}(E) \quad \text{(B5)}$$

and

$$D^<(\omega) = \begin{cases} -2\pi i[1+N(|\omega|)]\rho_{ph}(\omega) & \omega < 0 \\ -2\pi i N(\omega)\rho_{ph}(\omega) & \omega > 0 \end{cases}; \quad D^>(\omega) = \begin{cases} -2\pi i N(|\omega|)\rho_{ph}(\omega) & \omega < 0 \\ -2\pi i[1+N(\omega)]\rho_{ph}(\omega) & \omega > 0 \end{cases}$$
(B6)

This leads to an expression for $\Gamma_{ph}(E)$, the phonon contribution to the electronic linewidth, that was first derived in Ref. [32]

$$\begin{aligned}\Gamma_{ph}(E) = 2\pi M^2 \int_0^\infty d\omega\, \rho_{ph}(\omega)\bigl\{&[1+N(\omega)]\rho_{el}(E-\omega)[1-n(E-\omega)] \\ &+ N(\omega)\rho_{el}(E+\omega)[1-n(E+\omega)] + N(\omega)\rho_{el}(E-\omega)n(E-\omega) \\ &+ [1+N(\omega)]\rho_{el}(E+\omega)n(E+\omega)\bigr\}\end{aligned} \quad \text{(B7)}$$

In cases where the molecule-lead interaction stems from chemisorption it is reasonable to expect that $\Gamma \gg \Gamma_{ph}$, so $\Gamma_{tot}$ may be replaced by $\Gamma$ in the RHS of Eq. (B3). Using this in expression (B1) for the inelastic current leads to

$$I_{inel} = \frac{2e}{\hbar}\int_{-\infty}^{+\infty} dE\, \frac{\Gamma_L \Gamma_R}{\Gamma^2}\rho_{el}(E)\Gamma_{ph}(E)\bigl(f_L(E) - f_R(E)\bigr) \quad \text{(B8)}$$

Next consider the elastic current. It is convenient to redefine again the zero-order Green function so that 'zero order' includes the coupling to the contacts as well the level shift (real part of the self energy) arising from the electron-phonon interaction. With this definition we have

$$G_0^r(E) = \frac{1}{E - \tilde{E}_1(E) + i\Gamma/2}; \quad \tilde{E}_1(E) = E_1 + \text{Re}\Sigma_{ph}^r(E) \tag{B9}$$

and

$$G^r(E) = G_0^r(E) + G_0^r(E)\left[i\,\text{Im}\Sigma_{ph}^r(E)\right]G^r(E) \tag{B10}$$

From Eq. (B10) it follows that

$$G^r(E)G^a(E) = \frac{2\pi\rho_{el}^{(0)}(E)}{\Gamma}\left\{1 - 2\pi\rho_{el}(E)\frac{\Gamma_{ph}(E)}{2}\left(1 - \frac{\Gamma_{ph}(E)}{2\Gamma}\right)\right\} \tag{B11}$$

Using this in Eq. (28) and taking again $\Gamma_{ph}/\Gamma \ll 1$ leads to

$$I_{el} = I_{el}^{(0)} + \delta I_{el} \tag{B12a}$$

$$I_{el}^{(0)} = \frac{2e}{\hbar}\int_{-\infty}^{\infty} dE\,\frac{\Gamma_L\Gamma_R}{\Gamma}\rho_{el}^{(0)}\left(f_L(E) - f_R(E)\right) \tag{B12b}$$

$$\delta I_{el} = -2\pi\frac{2e}{\hbar}\int_{-\infty}^{+\infty} dE\,\frac{\Gamma_L\Gamma_R}{\Gamma}\rho_{el}^{(0)}\rho_{el}\frac{\Gamma_{ph}}{2}\left(f_L(E) - f_R(E)\right) \tag{B12c}$$

Eqs. (B8) and (B12c) finally give the total correction to the tunneling current

$$\delta I = \delta I_{el} + I_{inel} = \frac{2e}{\hbar}\int_{-\infty}^{+\infty} dE\,\frac{\Gamma_L\Gamma_R}{\Gamma^2}\rho_{el}(E)\Gamma_{ph}(E)\frac{\left(E - \tilde{E}_1(E)\right)^2 - (\Gamma/2)^2}{\left(E - \tilde{E}_1(E)\right)^2 + (\Gamma/2)^2}\left(f_L(E) - f_R(E)\right) \tag{B13}$$

which is Eq. (33). Eq. (B13) is of the same form as Eq. (17) of Ref. [31], however the densities of electronic states $\rho_{el}$ and the phonon contribution to the resonance width $\Gamma_{ph}$ appear here in their exact form (i.e. including corrections due to the electron-phonon coupling) rather then in the zero-order (in *M*) forms as in Ref. [31], and the shifted resonance energy $\tilde{E}_1$ replaces the bare energy $E_1$. This makes it possible to use (B13) in conjunction with the SCBA scheme to evaluate SCBA-corrected lineshapes and linewidths of IETS features.

For low temperature, the Fermi functions in (B13) can be replaced by the corresponding step-functions, e.g. $f_L(E) = \theta(\mu_L - E)$. Introducing the voltage division factor $\eta$ that determines the position of the resonance level $E_1$ relative to $\mu_L$ and $\mu_R$ according to

$$\begin{aligned} \mu_L &= E_F + \eta |e| \Phi \\ \mu_R &= E_F - (1-\eta)|e|\Phi \end{aligned} \quad (B14)$$

one gets from (B13) the phonon contribution to the differential conductance in the form

$$\frac{d(\delta I)}{d\Phi} = \frac{2e^2}{\hbar} \frac{\Gamma_L \Gamma_R}{\Gamma^2} \left\{ \eta \rho_{el}(\mu_L) \Gamma_{ph}(\mu_L) \frac{(\mu_L - \tilde{E}_1(\mu_L))^2 - (\Gamma/2)^2}{(\mu_L - \tilde{E}_1(\mu_L))^2 + (\Gamma/2)^2} \right.$$

$$\left. + (1-\eta)\rho_{el}(\mu_R)\Gamma_{ph}(\mu_R)\frac{(\mu_R - \tilde{E}_1(\mu_R))^2 - (\Gamma/2)^2}{(\mu_R - \tilde{E}_1(\mu_R))^2 + (\Gamma/2)^2} \right\} \quad (B15)$$

Also at low $T$ we can take $N(\omega)=0$ and $n(E) \approx \frac{\Gamma_L}{\Gamma}\theta(\mu_L - E) + \frac{\Gamma_R}{\Gamma}\theta(\mu_R - E)$. Eq. (B7) then yields for $\Phi > 0$

$$\Gamma_{ph}(\mu_L) = 2\pi M^2 \int_0^{|e|\Phi} d\omega \rho_{ph}(\omega) \frac{\Gamma_R}{\Gamma} \rho_{el}(\mu_L - \omega) \quad (B16a)$$

$$\Gamma_{ph}(\mu_R) = 2\pi M^2 \int_0^{|e|\Phi} d\omega \rho_{ph}(\omega) \frac{\Gamma_L}{\Gamma} \rho_{el}(\mu_R + \omega) \quad (B16b)$$

Since $\rho_{ph}(\omega)$ is strongly peaked about $\omega = \Omega_0$, these functions of $\Phi$ behave as step functions near $|e\Phi| = \Omega_0$. In taking the derivative of (B15) with respect to $\Phi$, the main contribution arises from this near-singularity of $d\Gamma_{ph}/d\Phi$, other contributions constituting a background. Thus we finally obtain

$$\frac{d^2 \delta I}{d\Phi^2} = \frac{4\pi e}{\hbar} \frac{\Gamma_L \Gamma_R}{\Gamma^2} M^2 \rho_{ph}(|e|\Phi) \left\{ \eta \frac{\Gamma_R}{\Gamma} \rho_{el}(\mu_L)\rho_{el}(\mu_L - |e|\Phi) \frac{(\mu_L - \tilde{E}_1(\mu_L))^2 - (\Gamma/2)^2}{(\mu_L - \tilde{E}_1(\mu_L))^2 + (\Gamma/2)^2} \right.$$

$$\left. + (1-\eta)\frac{\Gamma_L}{\Gamma} \rho_{el}(\mu_R)\rho_{el}(\mu_R + |e|\Phi) \frac{(\mu_R - \tilde{E}_1(\mu_R))^2 - (\Gamma/2)^2}{(\mu_R - \tilde{E}_1(\mu_R))^2 + (\Gamma/2)^2} \right\}$$

$$+ \quad background \quad terms$$

$$(B17)$$

**Acknowledgements:** MR thanks the DoD/MURI initiative, the NASA URETI program and the NSF NCN program for support. AN thanks the Israel Science Foundation, the US-Israel Binational Science Foundation and the Volkswagen Foundation for support. AN thanks Prof. Jan van Ruitenbeek (Leiden University) for illuminating discussions.

**Figure captions**

**Figure 1**. A schematic view of the level structure for inelastic electron tunneling. The shaded areas on the right and left denote the continuous manifolds of states of the two leads where the lines separating the occupied and unoccupied states are the corresponding Fermi energies. For the right lead two manifolds are shown: one where the corresponding molecular state is the ground vibrational state of the molecule, and the other (diagonally shaded) where the molecule is in the first excited vibrational state. The horizontal dotted lines at heights $\mu_L$ and $E_1$ are added to guide the eye.

**Figure 2.** (a) The current-voltage dependence in a junction characterized by a single bridge level and by the "standard" set of parameters (see text) with $\Gamma_L = \Gamma_R = 0.5$eV. Full line (red) – result obtained for $M=0$ (no electron-phonon coupling). Dashed and dotted lines show the effect of electron-phonon coupling, where the dotted (blue) line represents the result obtained under the assumption that the molecular vibration is at thermal equilibrium unaffected by its coupling to the non-equilibrium electronic system while the dashed (green) line corresponds to the case where relaxation rate of the molecular vibration due to its coupling to its thermal environment (secondary phonons) is finite. Integrations over the electronic energy axis were done using an energy grid of 3501 points spanning the region between –0.5 and 3 eV with step-size 0.001 eV and those over the phonon frequency were carried using a grid of 601 points spanning the frequency range between -0.3 and 0.3 eV with step-size 0.001 eV. (b) The power loss (dashed (green) line, left axis) and the non-

equilibrium oscillator 'temperature' (full (red) line, right axis) for the case represented by dashed line in (a) plotted against the voltage bias. The dotted (blue) line depicts the power loss for the case represented by the dotted (blue) line in (a). (c) Same as the dashed line in (b), where the power loss is displayed relative to the total available power $I\Phi$.

**Figure 3**. Same as Fig. 2, for a two electronic state bridge (see text for the 'standard' parameters), comparing the ballistic case and the case with electron-phonon coupling and finite thermal relaxation rate.

**Figure 4**. The inelastic tunneling spectrum, $d^2I/d\Phi^2$ plotted against the bias potential $\Phi$, for the model with a single bridge state at $T=10K$. The 'standard' parameters are used with $\Gamma_L=0.05eV$ $\Gamma_R=0.5eV$. The calculation is done using the SCBA (full line, red), the BA (dashed line, green) and the LOPT (dotted line, blue) schemes. These calculations were done using a grid in electron energy of 1501 points spanning range from -1.0 eV to 2.75 eV with steps size 0.0025 eV and a grid in the phonon frequency of 201 points spanning range from –0.5 eV to 0.5 eV with step 0.0025 eV.

**Figure 5**. The transmission coefficient T plotted as a function of electron energy $E$ for the junction of Fig. 2 where the electron escape rates $\Gamma_L$ and $\Gamma_R$ are taken each to be 0.013eV. Integration grids are the same as in figs. 2 and 3.

**Figure 6**. $d^2I/d\Phi^2$ plotted against $\Phi$ for the single resonant level model with parameters described in the text. Full line (red) – SCBA. Dashed line (green): BA Dotted line (blue): LOPT. The inset shows an expanded view of the LOPT result. The integration grids are the same as in Fig. 4.

**Figure 7**. The IETS threshold feature in $d^2I/d\Phi^2$ for the one resonant level model using the 'standard' parameters (see text) with $\Gamma_L=\Gamma_R=0.5eV$ and $T=10K$. Full line (red) – $E_1=0.70eV$, dashed line (green) – $E_1=0.60eV$, dotted line (blue) – $E_1=0.55eV$. The grids used in the numerical calculation are the same as in Fig. 4.

**Figure 8**. The IETS threshold feature in $d^2I/d\Phi^2$ for the one resonant level model at $T=10K$, using the 'standard' parameters (see text) except that $E_1$ is taken as 0.6eV. $\Gamma_R=0.5eV$ and $\Gamma_L$ is 0.05eV in (a) and 0.5eV in (b). In (b) the full line corresponds to the case where the Fermi energies are shifted under the bias according to $\mu_L = |e|\Phi(\Gamma_R/\Gamma)$, $\mu_R = -|e|\Phi(\Gamma_L/\Gamma)$, while the dashed line was produced for the model $\mu_L = |e|\Phi$, $\mu_R = 0$. Numerical grids are as in Fig. 6.

**Figure 9**. The width of a vibrational IETS feature obtained from Eq. (42) (with the additional additive term $\gamma_{ph} = 0.0001 eV$) plotted against the imposed potential bias. See text for parameters. Full (red) line – SCBA calculation. Dotted (blue) line – lowest order perturbation theory.

40